%% file: EuCAP2024_template.tex
\documentclass[conference,a4paper]{IEEEtran}
\usepackage{xcolor}
\usepackage{balance}
\usepackage[nolist]{acronym}
\usepackage[T1]{fontenc}

\ifCLASSINFOpdf
  \usepackage[pdftex]{graphicx}
\else
  \usepackage[dvips]{graphicx}
\fi
\ifCLASSOPTIONcompsoc
 \usepackage[caption=false,font=normalsize,labelfont=sf,textfont=sf]{subfig}
\else
 \usepackage[caption=false,font=footnotesize]{subfig}
\fi
\begin{document}

\input{acronyms.tex}
%
\title{K-factor Evaluation in a Hybrid Reverberation Chamber plus CATR OTA Testing Setup}

\author{\IEEEauthorblockN{
Alejandro Antón Ruiz\IEEEauthorrefmark{1},   
Samar Hosseinzadegan\IEEEauthorrefmark{2},   
John Kvarnstrand\IEEEauthorrefmark{2},    
Klas Arvidsson\IEEEauthorrefmark{2},      
Andrés Alayón Glazunov\IEEEauthorrefmark{1}\IEEEauthorrefmark{3} 
}                                     
\IEEEauthorblockA{\IEEEauthorrefmark{1}
University of Twente, Enschede, The Netherlands, \{a.antonruiz a.alayonglazunov\}@utwente.nl}
\IEEEauthorblockA{\IEEEauthorrefmark{2}
Bluetest AB, Gothenburg, Sweden, name.familyname@bluetest.se}
\IEEEauthorblockA{\IEEEauthorrefmark{3}
Linköping University, Norrköping Campus, Sweden, andres.alayon.glazunov@liu.se}
}



\maketitle

\begin{abstract}
This paper investigates achieving diverse K-factors using a Reverberation Chamber (RC) with a Compact Antenna Test Range (CATR) system. It explores six hybrid "RC plus CATR" configurations involving different excitations of the Rich Isotropic Multipath (RIMP) field and CATR-generated plane waves, with some setups including absorbers. A fixed horn antenna points towards the CATR in all configurations. The study found that the null hypothesis of Rayleigh or Rician probability distributions for the received signal envelope could not be rejected, with RIMP setups primarily conforming to Rayleigh distribution and all setups showing Rician distribution. Various K-factors were obtained, but no generalizable method for achieving the desired K-factor was identified. The paper also estimates the K-factor as a function of frequency in the 24.25-29.5 GHz band. Smaller K-factors exhibit larger fluctuations, while larger K-factors remain relatively stable, with consistent fluctuations across the frequency range.
\end{abstract}

\vskip0.5\baselineskip

%

\section{Introduction}

\ac{OTA} testing is key for developing and final product testing of wireless devices. It provides an assessment of real-life performance in controlled and repeatable conditions. \ac{OTA} testing is currently the only option for fully integrated devices without any connectors available and for smart antenna systems \cite{5G_Testing_Survey}. It is impractical to characterize the device in infinite scenarios to ensure it always performs well. In practice, it is necessary to constrain the testing to a subset of meaningful scenarios. A relevant parameter for a testing scenario is the distribution of the received signal at the \ac{DUT}. Two limiting propagation scenarios can be identified: \ac{RIMP} \cite{Kildal_hyp}, and pure \ac{LOS}, having an application in the \ac{Random-LOS} technique for \ac{OTA} testing \cite{RandomLOS}. \ac{RIMP} can be achieved in a \ac{RC}, where a distribution of the received signal close to theoretical Rayleigh is generated if the chamber is properly designed \cite{LTEBOOK}. Pure \ac{LOS} can be achieved in an \ac{AC} or also in a semi-anechoic as used in the \ac{EMC} tests. 

The above two \ac{OTA} environments offer excellent methods to recreate the ideal limiting propagation scenarios. However, real propagation environments are much more complex and may require combining the \ac{RIMP} and the \ac{CATR}, depending on the application, as provided by the Bluetest solution.  Thus, a \ac{LOS} component and other \ac{NLOS} components can be generated following a desired mixture. In this case, the distribution of the envelope of the received signal at the \ac{DUT} should be a Rician. This distribution is characterized by two parameters: the $K-$factor defined as the ratio of the \ac{LOS} over the \ac{NLOS} powers, or the ratio between the unstirred power over the stirred power and the average received power \cite{StirUnstir}. An ideal \ac{RIMP} scenario has a $K-$factor of 0 (or -$\infty$ dB), while pure \ac{LOS} would have a $K-$factor of $\infty$.

On another note, \ac{mmWave} has been deemed as one of the enablers for \ac{5G} communications \cite{mmWavejust} due to the large available bandwidth at those frequencies. Due to the smaller wavelengths, diffraction is less relevant in the propagation. Also, channels at \ac{mmWave} tend to be sparser, more directional, and have higher pathloss than sub-10 GHz frequencies. Therefore, it is expected that, in a general sense, $K-$factors of \ac{mmWave} channels are higher than those of lower frequencies. Highly directive antennas such as phased arrays are used to overcome pathloss.

In this work, an investigation of the use of different configurations of a \ac{RC} to achieve different $K-$factors is performed. The frequency range of interest is the lower FR2 bands for \ac{5G} ($24.25$~GHz to $29.5$~GHz). The \ac{RC} used for this experiment is a Bluetest RTS65, with the \ac{CATR} option installed \cite{BluetestCATR}. This allows to generate, in addition to the \ac{RIMP} and the \ac{CATR} environments separately, also other channels where both are mixed. In addition, using an absorber designed to attenuate the reflections coming from the \ac{CATR} has also been considered in this work. The emulation of Rician channels in \acp{RC} is not new in the literature \cite{RC_Rician,KFEmulRician, Emul_Rician_Andres_RC, MIMORician}. However, to the author's best knowledge, this investigation has only been carried out at sub-6 GHz frequencies. This work considers using a \ac{CATR} system at \ac{mmWave} frequencies. Existing commercial products are used novelly since the Bluetest RTS65 was not originally designed to emulate Rician channels.

\section{Signal Probability Distribution}

\subsection{Rayleigh Probability Distribution}
In a \ac{RIMP} environment, the distribution of the receive signal envelope by the \ac{DUT} positioned in the test zone is Rayleigh and results from multipath interference. The scale parameter $\sigma$ fully determines the distribution \cite{LTEBOOK}. In this case, the power follows an exponential distribution, with the parameter $\sigma^2$ as its only parameter, the average receive power. Estimating this parameter is straightforward by performing the mean of the obtained data.

\subsection{Rician Probability Distribution}
The Rician distribution model signals have two components: deterministic or direct \ac{LOS} and \ac{NLOS} random or scattered field components. The latter is Rayleigh distributed \cite{RC_Rician}. The $K-$factor is the power ratio of the \ac{LOS} component over the \ac{NLOS} component. In an ideal \ac{RC}, which produced a perfect Rayleigh distribution, the $K-$factor would be 0 (or $-\infty$ dB). However, that is not achievable in practice. In the \ac{RC} context, the $K-$factor can be better understood as
\begin{equation}
K=P_d/P_s,
\label{Eq1}
\end{equation}
where $P_s$ is the stirred power, and $P_d$ is the power of the unstirred paths, which includes a direct \ac{LOS} when the \ac{CATR} is used.

In this work, we have decided to use the estimator found in \cite{KFEmulRician} since it is unbiased. A major note is that this estimator requires acquiring the complex value of the transmission parameter $S_{21}$. While we do not reproduce the estimator equation, we say that it depends on the number of samples, which are assumed to be independent. This we take for granted in this work. However, future work will fully consider their independence and impact on the estimation.

\section{Experimental setup}
\subsection{Chamber}
\begin{figure}[!t]
\centering
\includegraphics[width=0.81\columnwidth]{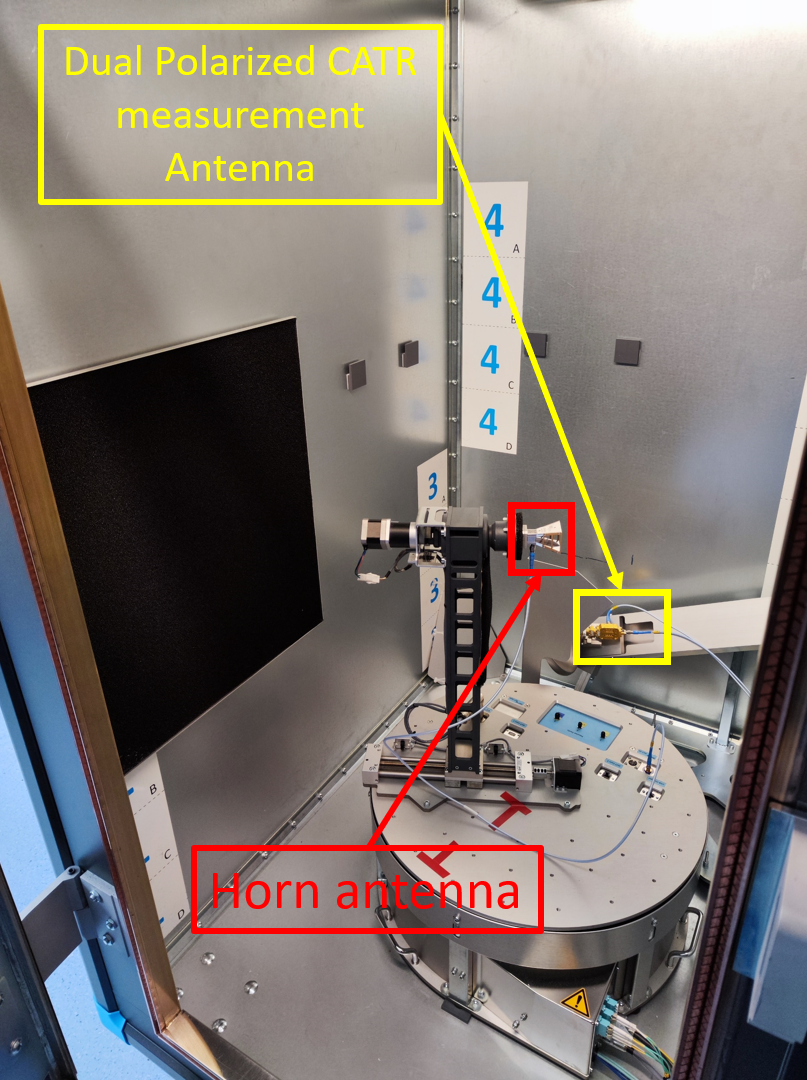}
\caption{Bluetest RTS65 with the \ac{CATR} option and the back absorber installed. The dimensions of the chamber are $1945$x$2000$x$1440$~mm (WxHxD).}
\label{F1}
\end{figure}

Fig.~\ref{F1} shows the experimental setup using a Bluetest RTS65 \ac{RC} chamber with the \ac{CATR} option installed \cite{BluetestCATR}. The chamber is designed to be used as a 2-in-1 test system, providing a \ac{RIMP} field in the test zone when it is excited through the \ac{RIMP} port or a \ac{LOS} only when it is excited through the \ac{CATR} port. The latter mode uses absorbers on the walls, enabling radiation pattern measurements within the same space.

\begin{figure}[!t]
\centering
\includegraphics[width=0.81\columnwidth]{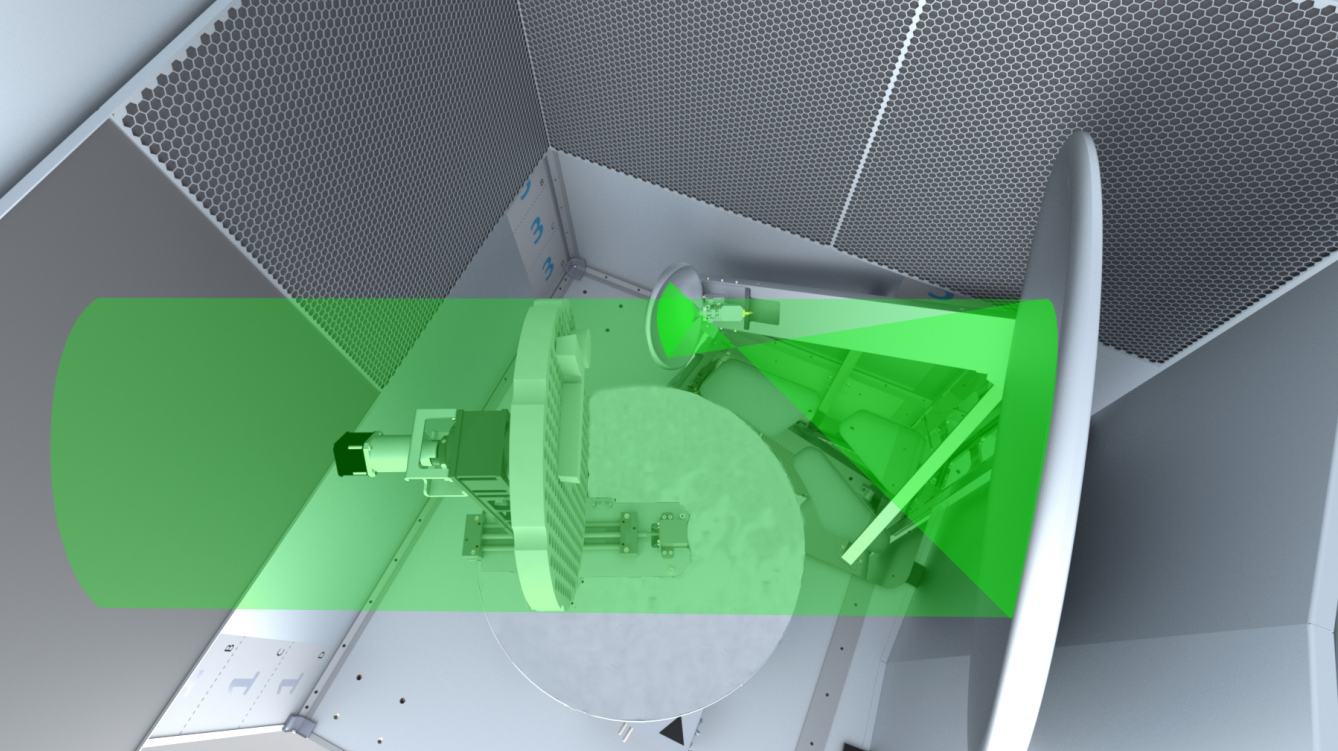}
\caption{CAD render showing geometry and signal flow (in green). The test zone is located within the lightest green volume. Source: \cite{BluetestCATR}.}
\label{F2}
\end{figure}
\begin{figure}[!t]
\centering
\includegraphics[width=0.81\columnwidth]{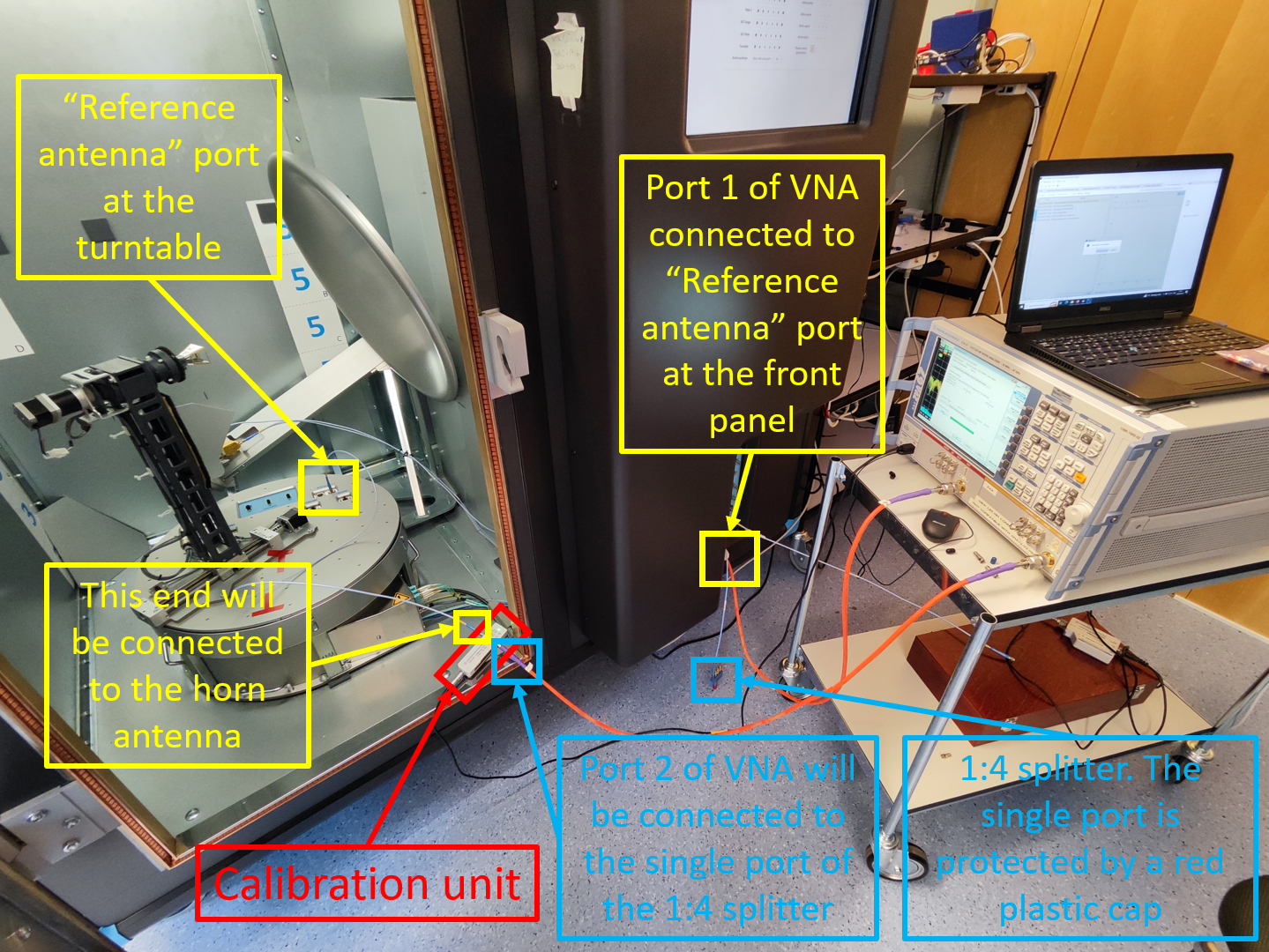}
\caption{Experiment setup for calibrating the \ac{VNA}.}
\label{F3}
\end{figure}

The \ac{CATR} mode of the chamber is realized by the use of a parabolic reflector with a dual-polarized antenna feed shown in yellow in Fig.~\ref{F1}. The generated plane wave has a $0.6$~dB amplitude and a $4^\circ$ phase ripples. The \ac{STD} ripples are at frequencies from $24.25-42$~GHz within a $30$~cm diameter cylindrical test zone \cite{CATR} (see Fig.~\ref{F2}). The polarization of the antenna is selected through a passive switch. To handle the reflections coming from the reflector or the \ac{DUT}, a set of carbon-loaded foam absorbers is placed around the chamber, as shown in Fig.~\ref{F2}. Except for the absorber on the wall towards which the plane wave is directed ("back absorber"), all the other absorbers are \ac{FSA}. They are covered by a metallic honeycomb pattern designed with spacing such that the sub-6 GHz frequency signals for which the chamber is also designed to operate in \ac{RIMP} mode are not significantly attenuated by the \ac{FSA}. However, they provide significant attenuation at \ac{mmWave}. The back absorber has the same material and size as the \ac{FSA}, except that the metallic honeycomb part is removed, thus providing better absorption. This is because most of the energy coming from the reflector is directed to this back absorber. In this work, only the back absorber scenario was considered (see Fig.~\ref{F1}).


\subsection{Antenna}

The experiment uses a linearly-polarized double ridged \ac{HA}, designed to operate from $4.5-50$~GHz \cite{DRH50}. The choice of antenna is because of its directional pattern and an accessible port so that the complex-valued $S_{21}$ can be measured to estimate the $K-$factor. The use of a directive antenna for \ac{mmWave} experiments is highly relevant because the obtained $K-$factor can serve as a reference value for other directive antennas. Even though results are valid for this specific antenna because of the spatial filtering that the directive antenna performs, they might be similar to those obtained for a different antenna with similar directivity. For example, results obtained with a phased array can be compared to the reference value. However, a characterization of the $K-$factor using an omnidirectional antenna is acknowledged to be relevant but is left for future work.

\subsection{Considered test cases}
A total of six cases are considered:

\begin{itemize}
    \item NoAs\_R: no absorbers, only RIMP excitation.
    \item NoAs\_C\_PS1: no absorbers, only CATR excitation with the matching polarization of the \ac{HA}.
    \item NoAs\_RC\_PS1: no absorbers, RIMP and CATR excitation with the CATR in the matching polarization of the \ac{HA}.
    \item BAs\_R: back absorber, only RIMP excitation.
    \item BAs\_C\_PS1: back absorber, only CATR excitation with the matching polarization of the \ac{HA}.
    \item BAs\_RC\_PS1: back absorber, RIMP and CATR excitation with the CATR in the matching polarization of the \ac{HA}.
\end{itemize}
Note that PS1 stands for "Passive Switch set to 1", which means that the \ac{CATR} port is routed to the port of the reflector feeder antenna designed as "1", which is the one that produces the polarization that matches that of the \ac{HA}, as it was installed in the experiment.

\subsection{Instrument and measurement}

\subsubsection{Cabling}

To perform this experiment, a 1:4 splitter was used to connect one of the \ac{VNA} ports to the \ac{RIMP} and/or \ac{CATR} ports, while the other port of the \ac{VNA} was connected to the \ac{HA} via a port in the front panel of the chamber. For \ac{RIMP} or \ac{CATR} only measurements, only one of the four splitter ports was used, while the other three were terminated. Two of the four ports were used in the \ac{RIMP} plus \ac{CATR} measurements, while the other two were terminated.

\subsubsection{VNA and measurement configuration}

The \ac{VNA} was configured to perform a $S_{21}$ sweep in the frequency from $24-29.5$~GHz with a $10$~MHz step, although only the $24.25$ to $29.5$~GHz range is presented here to adequate to the FR2 bands. An \ac{IF} bandwidth of $1$~KHz was used.

For each frequency point, a total of 600 samples were collected. Those 600 samples were taken at 600 unique positions of the two mode-stirrers of the chamber located on the ceiling and right wall while keeping the turntable fixed where the \ac{HA} was pointed at the reflector. It is worth noting in Fig.~\ref{F1} that the "right wall" is located on the right, behind the large reflector and the metallic plate behind it. Hence, keeping the antenna in a fixed position relative to the reflector was done to have a more controlled interaction between the \ac{HA} and the reflector. However, it is left for future work to use the turntable as a stirring mechanism, making the results less dependent on the direction towards which the main beam of the antenna is pointing.

Regarding the \ac{VNA} calibration, the calibration plane was, on one side, at the port of the \ac{HA} and, on the other side, at the single port of the 1:4 splitter. This is depicted in more detail in Fig~\ref{F3}.

\begin{figure*}
\centering
\subfloat[NoAs\_R]{\includegraphics[width=0.66\columnwidth]{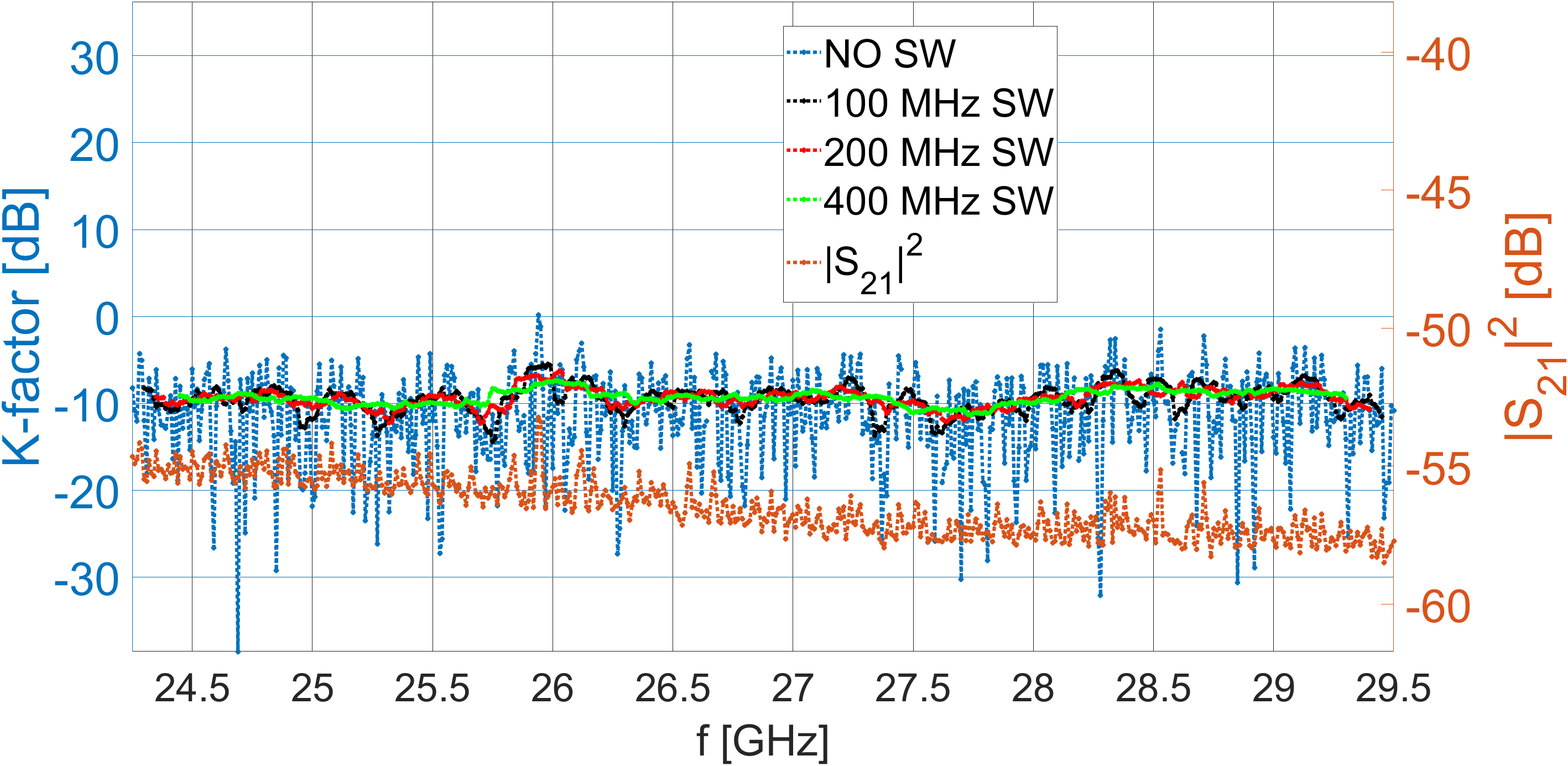}%
\label{F4a}}
\hfil
\subfloat[NoAs\_C\_PS1]{\includegraphics[width=0.66\columnwidth]{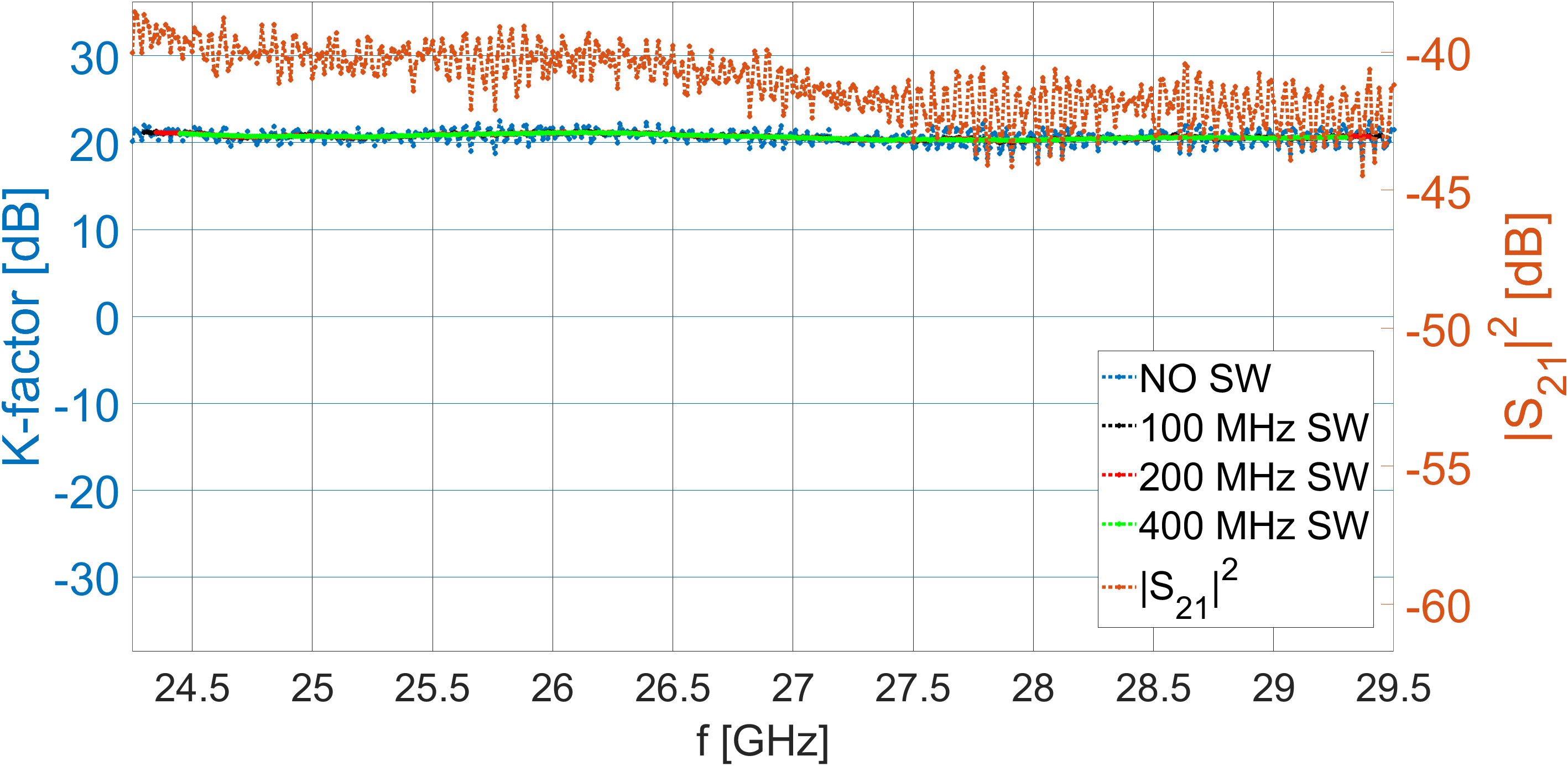}%
\label{F4b}}
\hfil
\subfloat[NoAs\_RC\_PS1]{\includegraphics[width=0.66\columnwidth]{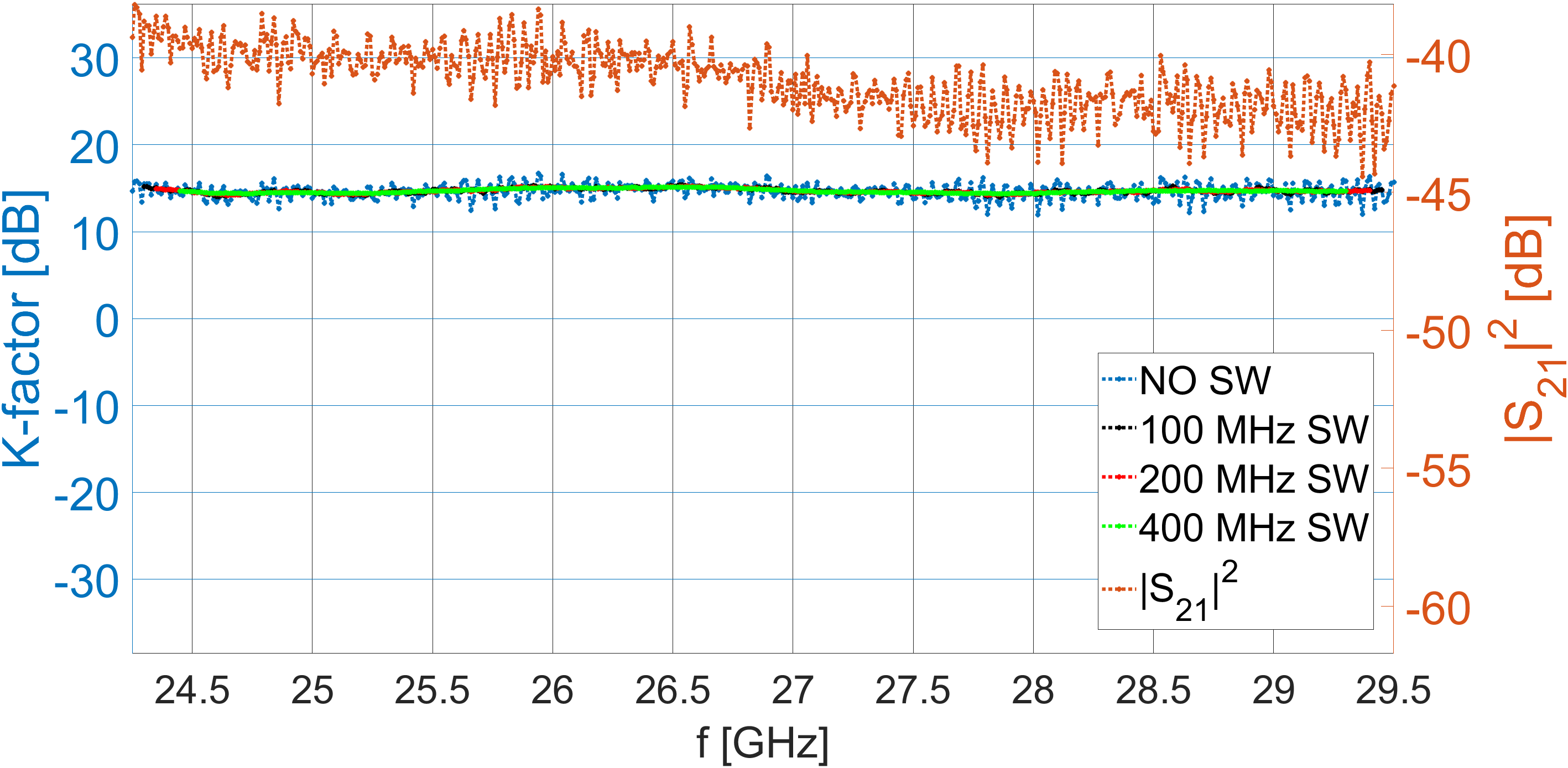}%
\label{F4c}}
\vfil
\subfloat[BAs\_R]{\includegraphics[width=0.66\columnwidth]{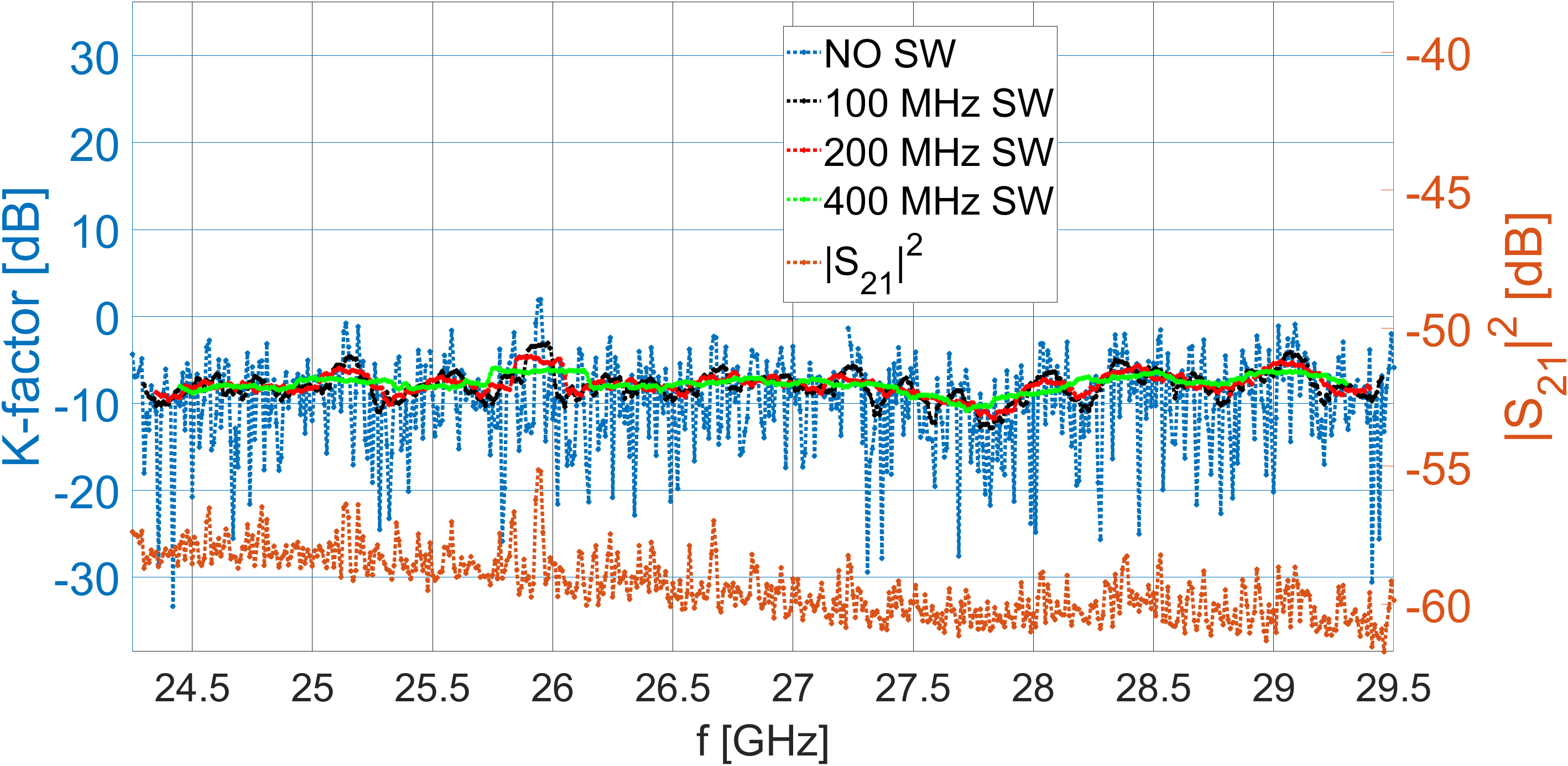}%
\label{F4d}}
\hfil
\subfloat[BAs\_C\_PS1]{\includegraphics[width=0.66\columnwidth]{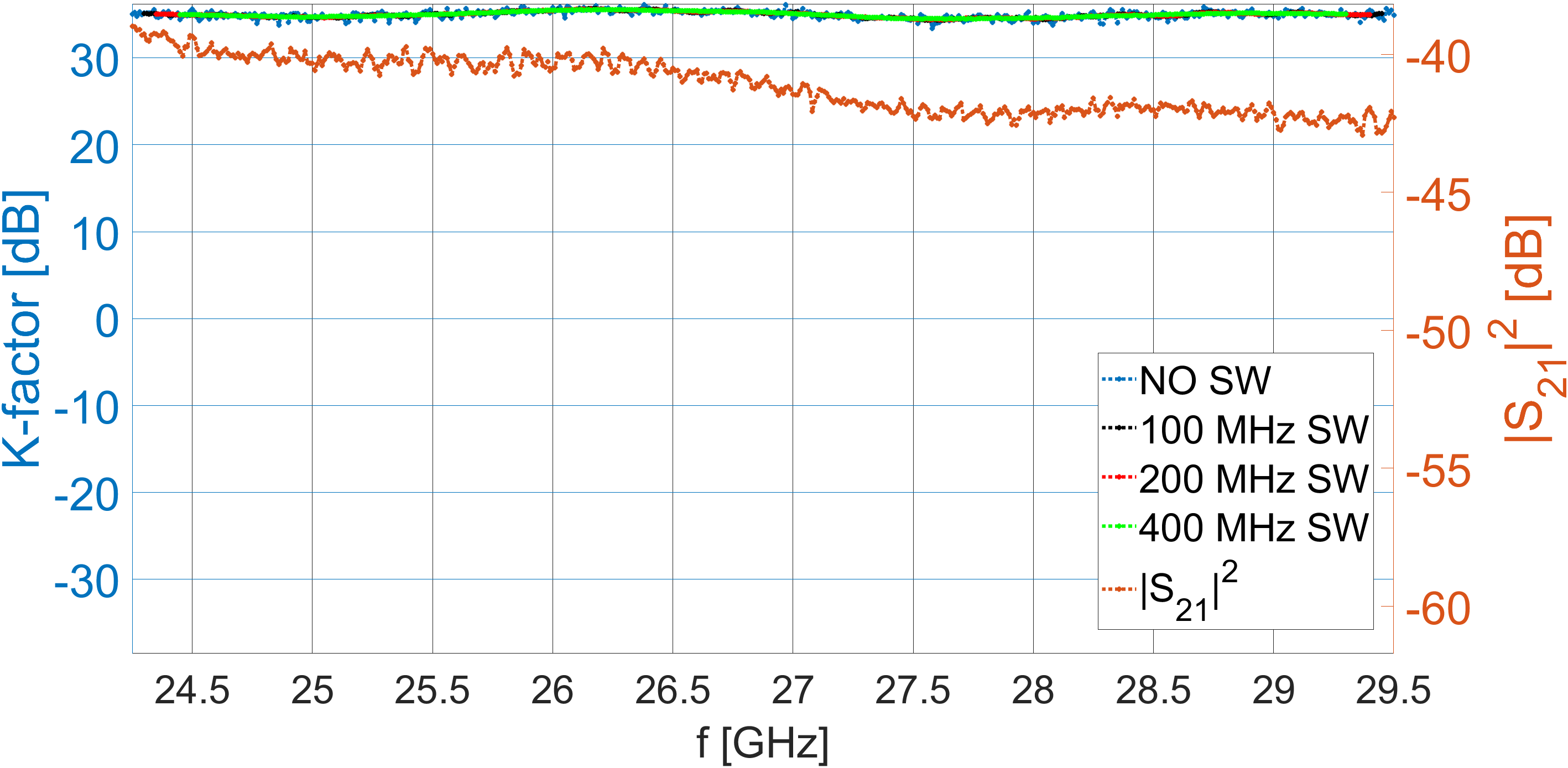}%
\label{F4e}}
\hfil
\subfloat[BAs\_RC\_PS1]{\includegraphics[width=0.66\columnwidth]{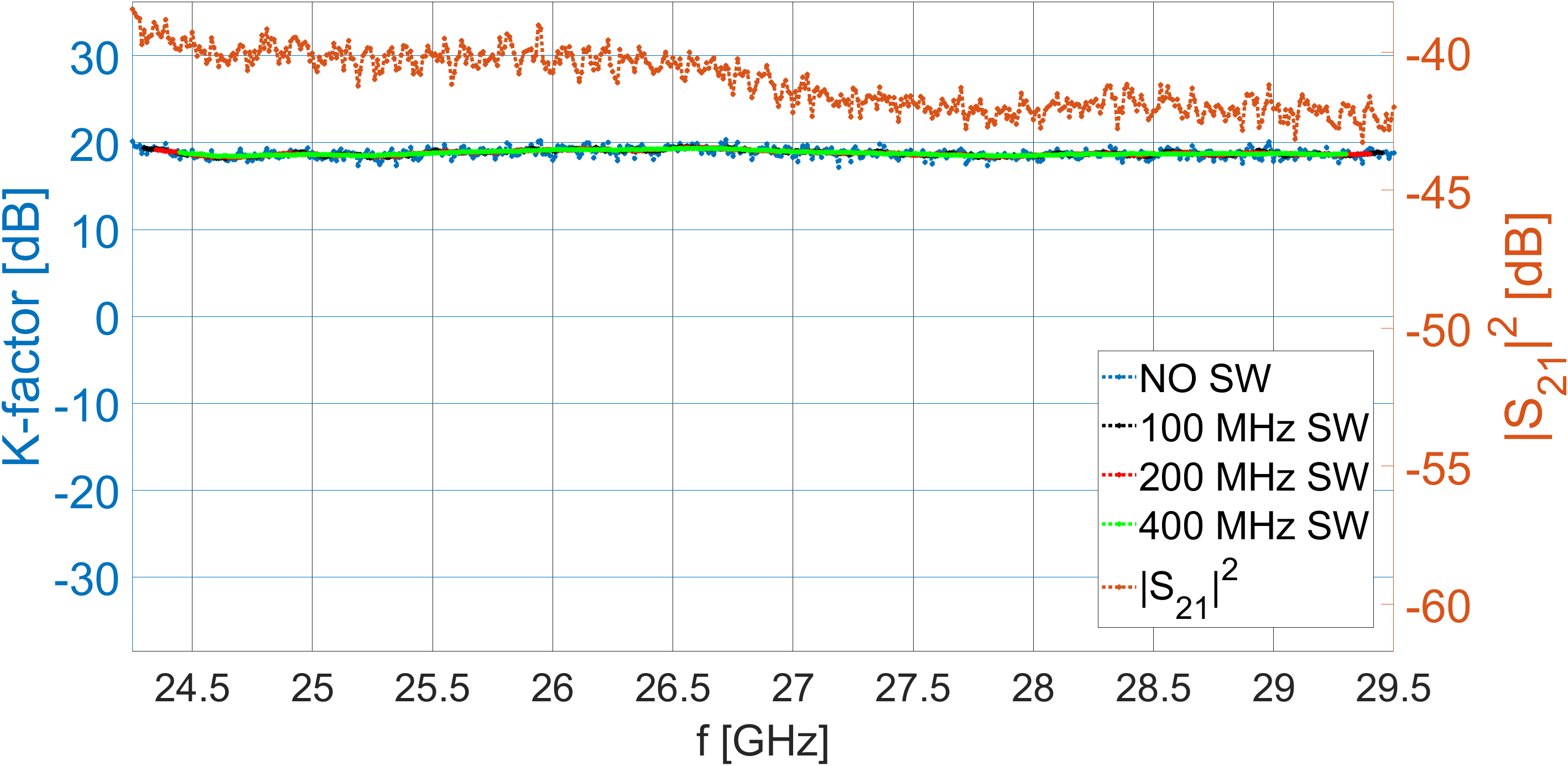}%
\label{F4f}}
\caption{$K-$factor for different \acp{SW} and $|S_{21}|^2$, as a function of the frequency; where only $|S_{21}|^2$ belongs to the right y-axis.}
\label{F4}
\end{figure*}

\begin{table*}
\centering
\caption{Relevant statistics of $K-$factor and $|S_{21}|^2$.}
\label{T1}
\includegraphics[width=1.5\columnwidth]{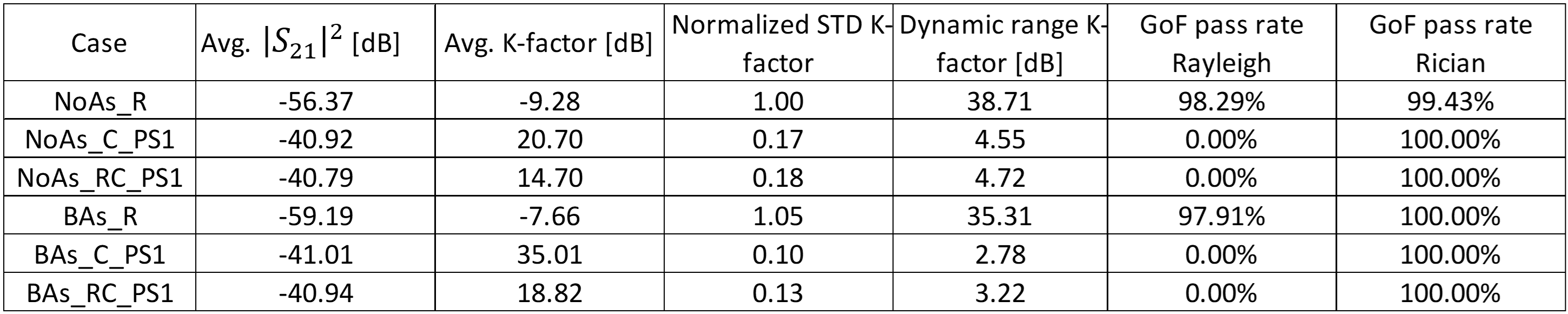}
\end{table*}

\section{Results}
Fig.~\ref{F4} shows the results of both $K-$factor (left y-axis) and $|S_{21}|^2$ (right y-axis) as a function of the considered frequency. The estimated $K-$factor is computed for each frequency point of the considered band and shown as “No SW,” where SW stands for Sliding Window. Then, an averaging of the $K-$factor over \acp{SW} of 100, 200, and 400 MHz is presented. This averaging is common in the literature to reduce statistical variations and is known as frequency stirring \cite{StirUnstir} and, in this case, it can approximate what the average $K-$factor value of a \ac{5G} FR2 channel of 100, 200, or 400 MHz centered at the x-axis frequency would be in this setup. The $|S_{21}|^2$ is plotted to assess what occurs in each studied case regarding the system losses. On the other hand, Table.~\ref{T1} shows some relevant statistics of both the $K-$factor and $|S_{21}|^2$, for the whole frequency range, including the results of the one-sample Kolmogorov-Smirnov tests that were performed to assess if the distributions of the 600 samples at each frequency could come from Rayleigh or Rician distributions. The presented result in Table~\ref{T1} is the pass rate of these \ac{GoF} tests, which indicates the percentage of frequencies for which it could not be rejected that the samples $|S_{21}|$ came from a fitted Rayleigh or Rician distribution, at a $5\%$ significance level. Regarding this, the results align with the expectations since for the \ac{RIMP} only excitations, the pass rate for Rayleigh is high, while, for the cases with \ac{CATR} excitation, it is 0. The pass rate for Rician distribution is high in all cases, aligned with the expectations, which enables the use of the $K-$factor as a relevant parameter for all cases. Note that the Rayleigh distribution is a particular case of the Rician distribution. For the \ac{CATR} only, even if we had a pure \ac{LOS} contribution, we would still have a particular case of Rician distribution (with $\infty$ $K-$factor).

In Fig.~\ref{F4}, it can be observed that the $K-$factor heavily fluctuates, in relative terms, i.e., in dB, whenever the chamber is not excited through the \ac{CATR} port (a, d), remaining much more stable when the \ac{CATR} port is excited (b, c, e, f). This can also be observed in Table~\ref{T1}, in both the dynamic range (or peak-to-peak variation) and the normalized \ac{STD} of the $K-$factor. The stability when the \ac{CATR} is excited is caused by the fact that the \ac{LOS} component, which is relatively stable with frequency, dominates over the stirred components. The large variations of $K-$factor when \ac{RIMP} is only excited are caused by several factors. On the one hand, there is the effect of limited sampling in conjunction with a low $K-$factor, as observed in Fig.~8 of \cite{KFEmulRician}. In short, the \acp{CI} (in dB scale) for the estimates of the $K-$factor, which can be found in (43) and (44) from \cite{KFEmulRician}, when $K-$factor is low, are broadened significantly (in dB scale). For reference, the $95\%$ \ac{CI} for 600 samples and a $K-$factor of $-9.28$~dB is $[-11.09,-8.01]$~dB, while, for a $K-$factor of $35.01$~dB, it is $[34.65,35.35]$~dB. However, the $K-$factor varies much further than those confidence intervals, so it is likely that the actual $K-$factor experiences large frequency variations when it is low. On the other hand, these variations could also be partially due to the absence of turntable stirring. This leaves the antenna in a fixed position, resulting in frequency-dependant unstirred paths at the antenna position. Therefore, it would be interesting to use turntable stirring and observe if variations are smoothed.   

When the back absorber is placed, there is an increase in the system's losses, i.e., average $|S_{21}|^2$. In particular, the total averaged $|S_{21}|^2$ decreases in $2.82$~dB for the \ac{RIMP} excitation only, $0.09$~dB for the \ac{CATR} excitation only, and $0.15$~dB for the \ac{RIMP} and \ac{CATR} excitation. As expected, the increase in the system's losses in relative terms, i.e., in dB, is lower for the cases where the \ac{CATR} is on and, therefore, the dominant contribution is that of the \ac{LOS}, which does not interact with the absorber. It can be observed that there is a direct relation between higher average $K-$factor and smaller increases in the system's losses, being the \ac{CATR} only case the one with the smallest increase, of just $0.09$~dB.

As for the use of \acp{SW}, they only make a difference in the \ac{RIMP} only cases since there is not such a large variation on the $K-$factor with the frequency for the rest of the considered cases. For the \ac{RIMP} only cases, the $100$~MHz \ac{SW} already provides a relevant smoothing, which improves with the larger \acp{SW}. However, even with the $400$~MHz \ac{SW}, there are some variations of the $K-$factor with the frequency. This is not observed when the \ac{CATR} is excited. This result is relevant because it implies that within the presented cases, the proposed setup can only provide a quite frequency-stable $K-$factor when the \ac{CATR} is excited, resulting in high $K-$factors. Conversely, for low $K-$factors, it cannot provide, in relative terms, a frequency-stable $K-$factor, which occurs when only the \ac{RIMP} branch is excited. As a point for future work, it is left to assess if a frequency-stable, low (at least lower than the ones obtained when the \ac{CATR} is excited) $K-$factor can be achieved by, e.g., exciting the \ac{CATR} and \ac{RIMP}, but attenuating the \ac{CATR} branch, or using turntable stirring. On another note, it is also worth noting that the $K-$factor variations do not follow any clear trend with respect to the frequency, i.e., there is not an appreciable linear dependency of neither the \ac{SW} averaged $K-$factor nor the $K-$factor variation.

Finally, when observing the full frequency band averaged $K-$factor from Table~\ref{T1}, it is worth noting that the fact of adding the back absorber increases the $K-$factor in around $14.31$~dB when only the \ac{CATR} is excited, but only in around $4.12$~dB when both the \ac{RIMP} and \ac{CATR} are excited. Therefore, the estimated stirred power decreases in $14.36$~dB for the \ac{CATR} and $4.18$~dB for the \ac{RIMP} and \ac{CATR} case, when adding the back absorber. As for the estimated unstirred power, it decreases in $0.06$~dB for both the \ac{CATR} only and the \ac{RIMP} and \ac{CATR} cases, when adding the back absorber. Therefore, it is clear that the addition of the back absorber has a greater effect on reducing the stirred power coming from the \ac{CATR} by capturing a significant amount of the energy that is not directly coupled to the horn antenna. This leads to a greater increase of $K-$factor for the \ac{CATR} only case than for the \ac{RIMP} and \ac{CATR} one. As for the \ac{RIMP} only case, adding the back absorber decreases the $K-$factor of $1.62$~dB. This comes from a decrease of the estimated stirred power of $3.03$~dB and a decrease of the estimated unstirred power of $1.41$~dB. All in all, adding the back absorber for \ac{RIMP} excitation only decreases the resemblance of the environment to a pure \ac{RIMP} one.  

\balance

%
%



\section{Conclusion}
This paper presents six \ac{RC} setups equipped with a \ac{CATR} system operating at the $24.25-29.5$~GHz band, analyzing field distribution using one-sample Kolmogorov-Smirnov \ac{GoF} tests and focusing on $K-$factor. The \ac{GoF} tests align with expectations, confirming a Rician distribution for all cases and a Rayleigh distribution for \ac{RIMP}-only excitation. Different $K-$factors are achievable with varying setups. Low estimated $K-$factors are less stable with frequency relative to high $K-$factors. $K-$factor variations and sliding window averages show no frequency dependence. The directional horn antenna used in this study aims to provide a reference for directive antennas in spatially selective channels. However, further work can improve predictability by incorporating turntable stirring and omnidirectional antennas. This will also allow us to characterize the $K-$factor independently from the orientation of the antenna, at least on the 2D horizontal plane. Other setups should be explored, such as attenuators in \ac{RIMP} and/or \ac{CATR} configurations. This work marks an initial step toward utilizing a commercialized \ac{mmWave} product for novel \ac{OTA} testing with a controllable $K-$factor.


\section*{Acknowledgment}
The work of Alejandro Antón is supported by the European Union’s Horizon 2020 Marie Skłodowska-Curie grant agreement No. 955629. Andrés Alayón Glazunov also kindly acknowledges funding from the ELLIIT strategic research environment (https://elliit.se/).



%

\bibliographystyle{IEEEtran}

\bibliography{References}

\end{document}

%% file: acronyms.tex
\begin{acronym}

\acro{2D}{Two Dimensions}%
\acro{2G}{Second Generation}%
\acro{3D}{Three Dimensions}%
\acro{3G}{Third Generation}%
\acro{3GPP}{Third Generation Partnership Project}%
\acro{3GPP2}{Third Generation Partnership Project 2}%
\acro{4G}{Fourth Generation}%
\acro{5G}{Fifth Generation}%

\acro{AI}{Artificial Intelligence}%
\acro{AoA}{Angle of Arrival}%
\acro{AoD}{Angle of Departure}%
\acro{AR}{Augmented Reality}%
\acro{AP}{Access Point}
\acro{AE}{Antenna Element}
\acro{AC}{Anechoic Chamber}
\acro{AUT}{Antenna Under Test}

\acro{BER}{Bit Error Rate}%
\acro{BPSK}{Binary Phase-Shift Keying}%
\acro{BRDF}{ Bidirectional Reflectance Distribution Function}%
\acro{BS}{Base Station}%

\acro{CA}{Carrier Aggregation}%
\acro{CDF}{Cumulative Distribution Function}%
\acro{CDM}{Code Division Multiplexing}%
\acro{CDMA}{Code Division Multiple Access}%
\acro{CPU} {Central Processing Unit}
\acro{CUDA}{Compute Unified Device Architecture}
\acro{CDF}{Cumulative Distribution Function}
\acro{CI}{Confidence Interval}
\acro{CVRP}{Constrained-View Radiated Power}
\acro{CATR}{Compact Antenna Test Range}
 
\acro{D2D}{Device-to-Device}%
\acro{DL}{Down Link}%
\acro{DS}{Delay Spread}%
\acro{DAS}{Distributed Antenna System}
\acro{DKED}{double knife-edge diffraction}
\acro{DUT}{Device Under Test}


\acro{EDGE}{Enhanced Data rates for GSM Evolution}%
\acro{EIRP}{Equivalent Isotropic Radiated Power}%
\acro{eMBB}{Enhanced Mobile Broadband}%
\acro{eNodeB}{evolved Node B}%
\acro{ETSI}{European Telecommunications Standards Institute}%
\acro{ER}{Effective Roughness}%
\acro{E-UTRA}{Evolved UMTS Terrestrial Radio Access}%
\acro{E-UTRAN}{Evolved UMTS Terrestrial Radio Access Network}%
\acro{EF}{Electric Field}
\acro{EMC}{Electromagnetic Compatibility}

\acro{FDD}{Frequency Division Duplexing}%
\acro{FDM}{Frequency Division Multiplexing}%
\acro{FDMA}{Frequency Division Multiple Access}%
\acro{FoM}{Figure of Merit}
\acro{FoV}{Field of View}
\acro{FSA}{Frequency Selective Absorber}
\acro{GI}{Global Illumination} %
\acro{GIS}{Geographic Information System}%
\acro{GO}{Geometrical Optics} %
\acro{GPU}{Graphics Processing Unit}%
\acro{GPGPU}{General Purpose Graphics Processing Unit}%
\acro{GPRS}{General Packet Radio Service}%
\acro{GSM}{Global System for Mobile Communication}%
\acro{GNSS}{Global Navigation Satellite System}%
\acro{GoF}{Goodness of Fit}
\acro{H2D}{Human-to-Device}%
\acro{H2H}{Human-to-Human}%
\acro{HDRP}{High Definition Render Pipeline}
\acro{HSDPA}{High Speed Downlink Packet Access}
\acro{HSPA}{High Speed Packet Access}%
\acro{HSPA+}{High Speed Packet Access Evolution}%
\acro{HSUPA}{High Speed Uplink Packet Access}
\acro{HPBW}{Half-Power Beamwidth}
\acro{HA}{Horn Antenna}

\acro{IEEE}{Institute of Electrical and Electronic Engineers}%
\acro{InH}{Indoor Hotspot} %
\acro{IMT} {International Mobile Telecommunications}%
\acro{IMT-2000}{\ac{IMT} 2000}%
\acro{IMT-2020}{\ac{IMT} 2020}%
\acro{IMT-Advanced}{\ac{IMT} Advanced}%
\acro{IoT}{Internet of Things}%
\acro{IP}{Internet Protocol}%
\acro{ITU}{International Telecommunications Union}%
\acro{ITU-R}{\ac{ITU} Radiocommunications Sector}%
\acro{IS-95}{Interim Standard 95}%
\acro{IES}{Inter-Element Spacing}
\acro{IF}{Intermediate Frequency}


\acro{KPI}{Key Performance Indicator}%
\acro{K-S}{Kolmogorov-Smirnov}

\acro{LB} {Light Bounce}
\acro{LIM}{Light Intensity Model}%
\acro{LOS}{Line-Of-Sight}%
\acro{LTE}{Long Term Evolution}%
\acro{LTE-Advanced}{\ac{LTE} Advanced}%
\acro{LSCP}{Lean System Control Plane}%
\acro{LSI} {Light Source Intensity}

\acro{M2M}{Machine-to-Machine}%
\acro{MatSIM}{Multi Agent Transport Simulation}
\acro{METIS}{Mobile and wireless communications Enablers for Twenty-twenty Information Society}%
\acro{METIS-II}{Mobile and wireless communications Enablers for Twenty-twenty Information Society II}%
\acro{MIMO}{Mul\-ti\-ple-In\-put Mul\-ti\-ple-Out\-put}
\acro{mMIMO}{massive MIMO}%
\acro{mMTC}{massive Machine Type Communications}%
\acro{mmW}{millimeter-wave}%
\acro{MU-MIMO}{Multi-User MIMO}
\acro{MMF}{Max-Min Fairness}
\acro{MKED}{Multiple Knife-Edge Diffraction}
\acro{MF}{Matched Filter}
\acro{mmWave}{Millimeter Wave}

\acro{NFV}{Network Functions Virtualization}%
\acro{NLOS}{Non-Line-Of-Sight}%
\acro{NR}{New Radio}%
\acro{NRT}{Non Real Time}%
\acro{NYU}{New York University}%
\acro{N75PRP}{Near-75-degrees Partial Radiated Power}%
\acro{NHPRP}{Near-Horizon Partial Radiated Power}%

\acro{O2I}{Outdoor to Indoor}%
\acro{O2O}{Outdoor to Outdoor}%
\acro{OFDM}{Orthogonal Frequency Division Multiplexing}%
\acro{OFDMA}{Or\-tho\-go\-nal Fre\-quen\-cy Di\-vi\-sion Mul\-ti\-ple Access}
\acro{OtoI}{Outdoor to Indoor}%
\acro{OTA}{Over-The-Air}

\acro{PDF}{Probability Distribution Function}
\acro{PDP}{Power Delay Profile}
\acro{PHY}{Physical}%
\acro{PLE}{Path Loss Exponent}
\acro{PRP}{Partial Radiated Power}

\acro{QAM}{Quadrature Amplitude Modulation}%
\acro{QoS}{Quality of Service}%

\acro{RCSP}{Receive Signal Code Power}
\acro{RAN}{Radio Access Network}%
\acro{RAT}{Radio Access Technology}%

\acro{RAN}{Radio Access Network}%
\acro{RMa}{Rural Macro-cell}%
\acro{RMSE} {Root Mean Square Error}
\acro{RSCP}{Receive Signal Code Power}%
\acro{RT}{Ray Tracing}
\acro{RX}{receiver}
\acro{RMS}{Root Mean Square}
\acro{Random-LOS}{Random Line-Of-Sight}
\acro{RF}{Radio Frequency}
\acro{RC}{Reverberation Chamber}
\acro{RIMP}{Rich Isotropic Multipath}

\acro{SB} {Shadow Bias}
\acro{SC}{small cell}
\acro{SDN}{Software-Defined Networking}%
\acro{SGE}{Serious Game Engineering}%
\acro{SF}{Shadow Fading}%
\acro{SIMO}{Single Input Multiple Output}%
\acro{SINR}{Signal to Interference plus Noise Ratio}
\acro{SISO}{Single Input Single Output}%
\acro{SMa}{Suburban Macro-cell}%
\acro{SNR}{Signal to Noise Ratio}
\acro{SU}{Single User}%
\acro{SUMO}{Simulation of Urban Mobility}
\acro{SS} {Shadow Strength}
\acro{STD}{Standard Deviation}
\acro{SW} {Sliding Window}


\acro{TDD}{Time Division Duplexing}%
\acro{TDM}{Time Division Multiplexing}%
\acro{TD-CDMA}{Time Division Code Division Multiple Access}%
\acro{TDMA}{Time Division Multiple Access}%
\acro{TX}{transmitter}
\acro{TZ}{Test Zone}
\acro{TRP}{Total Radiated Power}


\acro{UAV}{Unmanned Aerial Vehicle}%
\acro{UE}{User Equipment}%
\acro{UI}{User Interface}
\acro{UHD}{Ultra High Definition}
\acro{UL}{Uplink}%
\acro{UMa}{Urban Macro-cell}%
\acro{UMi}{Urban Micro-cell}%
\acro{uMTC}{ultra-reliable Machine Type Communications}%
\acro{UMTS}{Universal Mobile Telecommunications System}%
\acro{UPM}{Unity Package Manager}
\acro{UTD}{Uniform Theory of Diffraction} %
\acro{UTRA}{{UMTS} Terrestrial Radio Access}%
\acro{UTRAN}{{UMTS} Terrestrial Radio Access Network}%
\acro{URLLC}{Ultra-Reliable and Low Latency Communications}%
\acro{UHRP}{Upper Hemisphere Radiated Power}%

\acro{V2V}{Vehicle-to-Vehicle}%
\acro{V2X}{Vehicle-to-Everything}%
\acro{VP}{Visualization Platform}%
\acro{VR}{Virtual Reality}%
\acro{VNA}{Vector Network Analyzer}
\acro{VIL}{Vehicle-in-the-loop}

\acro{WCDMA}{Wideband Code Division Multiple Access}%
\acro{WINNER}{Wireless World Initiative New Radio}%
\acro{WINNER+}{Wireless World Initiative New Radio +}%
\acro{WiMAX}{Worldwide Interoperability for Microwave Access}%
\acro{WRC}{World Radiocommunication Conference}%

\acro{xMBB}{extreme Mobile Broadband}%

\acro{ZF}{Zero Forcing}

\end{acronym}